\documentclass[aps,prl,reprint,superscriptaddress,showpacs,floatfix]{revtex4-1}

\usepackage{graphicx}
\usepackage{dcolumn}
\usepackage{bm}
\usepackage[mathlines]{lineno}

\begin{document}

\title{Spontaneous graphitization of ultrathin cubic structures:
       A computational study}

\author{Pavel B. Sorokin}
\affiliation{Physics and Astronomy Department,
             Michigan State University,
             East Lansing, Michigan 48824, USA}
\affiliation{FSBI Technological Institute for
             Superhard and Novel Carbon Materials,
             Troitsk, Moscow, Russia 142190}
\affiliation{Moscow Institute of Physics and Technology,
             Dolgoprudny, Russia 141700}

\author{Alexander G. Kvashnin}
\affiliation{FSBI Technological Institute for
             Superhard and Novel Carbon Materials,
             Troitsk, Moscow, Russia 142190}
\affiliation{Moscow Institute of Physics and Technology,
             Dolgoprudny, Russia 141700}

\author{Zhen Zhu}
\affiliation{Physics and Astronomy Department,
             Michigan State University,
             East Lansing, Michigan 48824, USA}

\author{David Tom\'{a}nek}
    \email
    {tomanek@pa.msu.edu}%
\affiliation{Physics and Astronomy Department,
             Michigan State University,
             East Lansing, Michigan 48824, USA}

\date{\today} 

\begin{abstract}
Results based on {\em ab initio} density functional calculations
indicate a general graphitization tendency in ultrathin slabs of
cubic diamond, boron nitride, and many other cubic structures
including rocksalt. Whereas such compounds often show an energy
preference for cubic rather than layered atomic arrangements in
the bulk, the surface energy of layered systems is commonly lower
than that of their cubic counterparts. We determine the critical
slab thickness for a range of systems, below which a spontaneous
conversion from a cubic to a layered graphitic structure occurs,
driven by surface energy reduction in surface-dominated
structures.
\end{abstract}




\maketitle



Structural changes at surfaces including atomic relaxation and
reconstruction are a manifestation of the driving force
to minimize their total free energy\cite{{Somorjai},{Zangwill}}.
Atomic rearrangements are typically moderate at surfaces of
semi-infinite systems and in thick slabs so that the energy
penalty associated with structural mismatch at the interface
between the reconstructed surface and the unreconstructed bulk may
be limited. In ultra-thin slabs, surface contribution dominates
the total energy, as only a small fraction of atoms experience
bulk-like atomic environment. There, a large-scale reconstruction
involving not only the topmost layers, but the entire system may
yield the most stable structure. Examples of such large-scale
atomic rearrangements, driven by a tendency to reduce the surface
energy, are the observed graphitization of diamond
nanoparticles\cite{BanhartRPP99} and
nanowires\cite{diamond_nanowire_graphitization}, as well as
ultrathin SiC\cite{2D-SiC} and ZnO\cite{2D-ZnO} films. This
graphitization scenario, if energetically viable for a large range
of compounds, may turn into a valuable bottom-up approach to
synthesize ultrathin layered structures for nanoelectronics
applications in the post-graphene era.


We present results of {\em ab initio} density functional
calculations, which indicate a general graphitization tendency in
ultrathin slabs of cubic compounds. We find that an energy
preference for layered honeycomb rather than cubic structures in
ultrathin slabs is rather common, extending from
diamond\cite{Kvashnin14} and boron nitride to less obvious cubic
structures including silicon carbide, boron phosphide and
rocksalt. Whereas the bulk of such compounds shows an energy
preference for cubic rather than layered atomic arrangements, the
surface energy of systems with honeycomb layers is commonly lower
than that of their cubic counterparts. Whether the type of crystal
bonding is purely covalent, purely ionic, or a combination of the
two, the optimum structure of a slab results from an energy
competition between the energy preference for a honeycomb
structure at the surface and for a cubic atomic arrangement in the
bulk. We determine the critical slab thickness for a range of
systems, below which a spontaneous conversion from a cubic to a
layered graphitic structure occurs, driven by surface energy
reduction in surface-dominated structures.

\begin{figure}[tb]
\includegraphics[width=1.0\columnwidth]{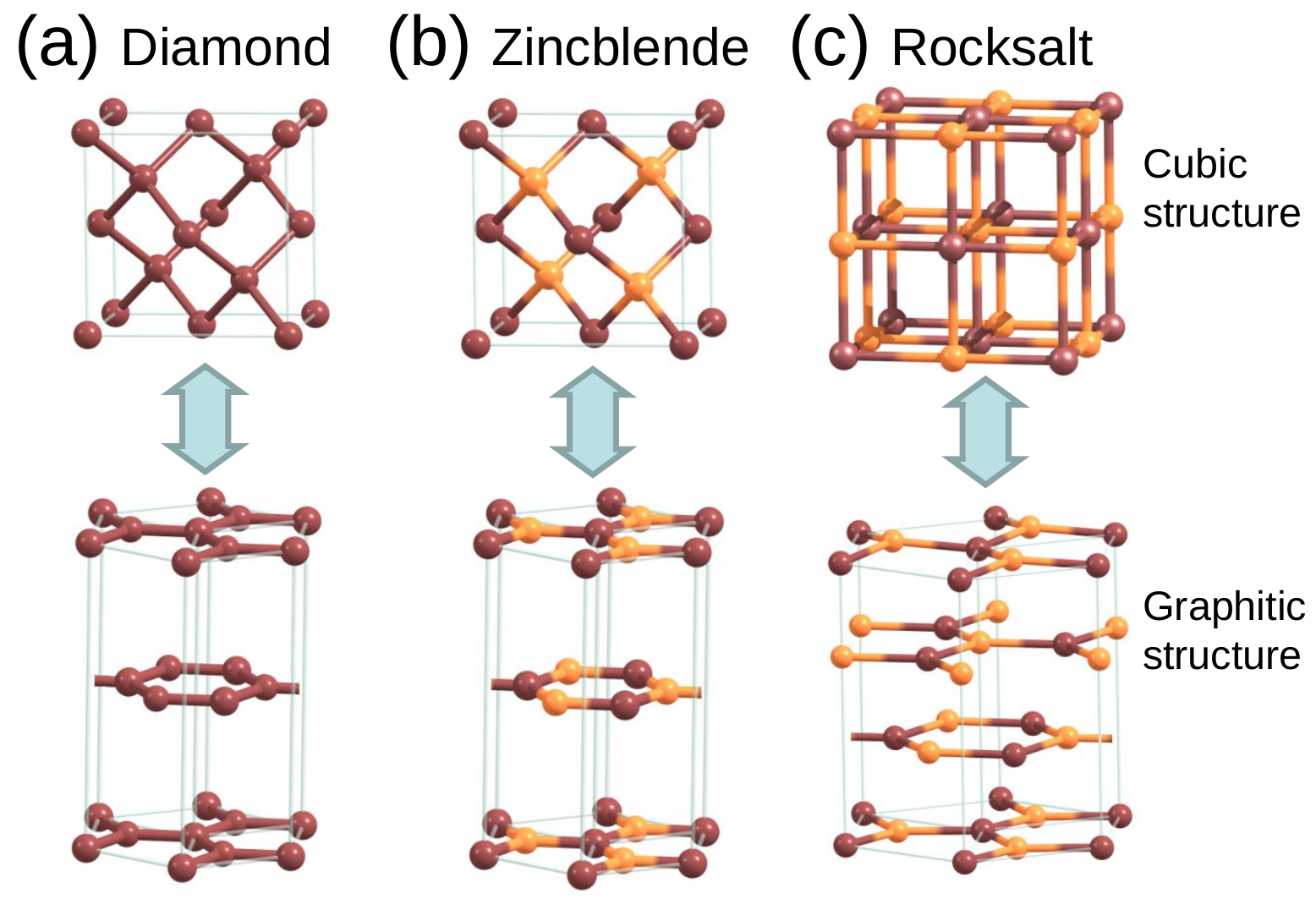}
\caption{(Color online) Ball-and-stick models of (a) diamond, (b)
zincblende, and (c) rocksalt in their native bulk structure (top
panels) and their corresponding layered counterparts (bottom
panels). \label{fig1}}
\end{figure}


Our computational approach to learn about the equilibrium
structure, stability and charge distribution in ultrathin slabs is
based on {\em ab initio} density functional theory (DFT) as
implemented in the \textsc{SIESTA}\cite{SIESTA} and
\textsc{VASP}\cite{VASP} codes. We used periodic boundary
conditions throughout the study, with multilayer structures
represented by a periodic array of slabs separated by a 15~{\AA}
thick vacuum region. We used the Perdew-Burke-Ernzerhof~\cite{PBE}
exchange-correlation functional throughout the study.
\textsc{VASP} calculations are based on the projector-augmented
wave method and our \textsc{SIESTA} studies make use of
norm-conserving Troullier-Martins
pseudopotentials~\cite{Troullier91} and a double-$\zeta$ basis
including polarization orbitals. The plane-wave energy cutoff was
set to 180~Ry in \textsc{SIESTA} and 500~eV in \textsc{VASP}. The
reciprocal space was sampled by a fine k-point
mesh\cite{Monkhorst-Pack76} ranging between $8{\times}8{\times}3$
and $6{\times}6{\times}1$ $k$-points in the Brillouin zone of the
primitive unit cell. All geometries have been optimized using the
conjugate gradient method\cite{CGmethod}, until none of the
residual Hellmann-Feynman forces exceeded $10^{-2}$~eV/{\AA}.


The inspiration for our study came from structure optimization
calculations for ultrathin slabs, constrained to a quasi-2D
geometry\cite{2D-constraints}, which indicated a spontaneous
transformation from a cubic to a layered graphitic structure in
systems ranging from diamond to rocksalt\cite{SM-cubgra14}. The
competing cubic and layered graphitic phases for diamond,
zincblende and rocksalt lattices are illustrated in
Fig.~\ref{fig1}. In the following, we investigate the
graphitization tendency of ultrathin slabs of such systems,
introduce a simple criterion to judge this tendency, and relate it
to the charge redistribution at surfaces.

\begin{figure}[tb]
\includegraphics[width=1.0\columnwidth]{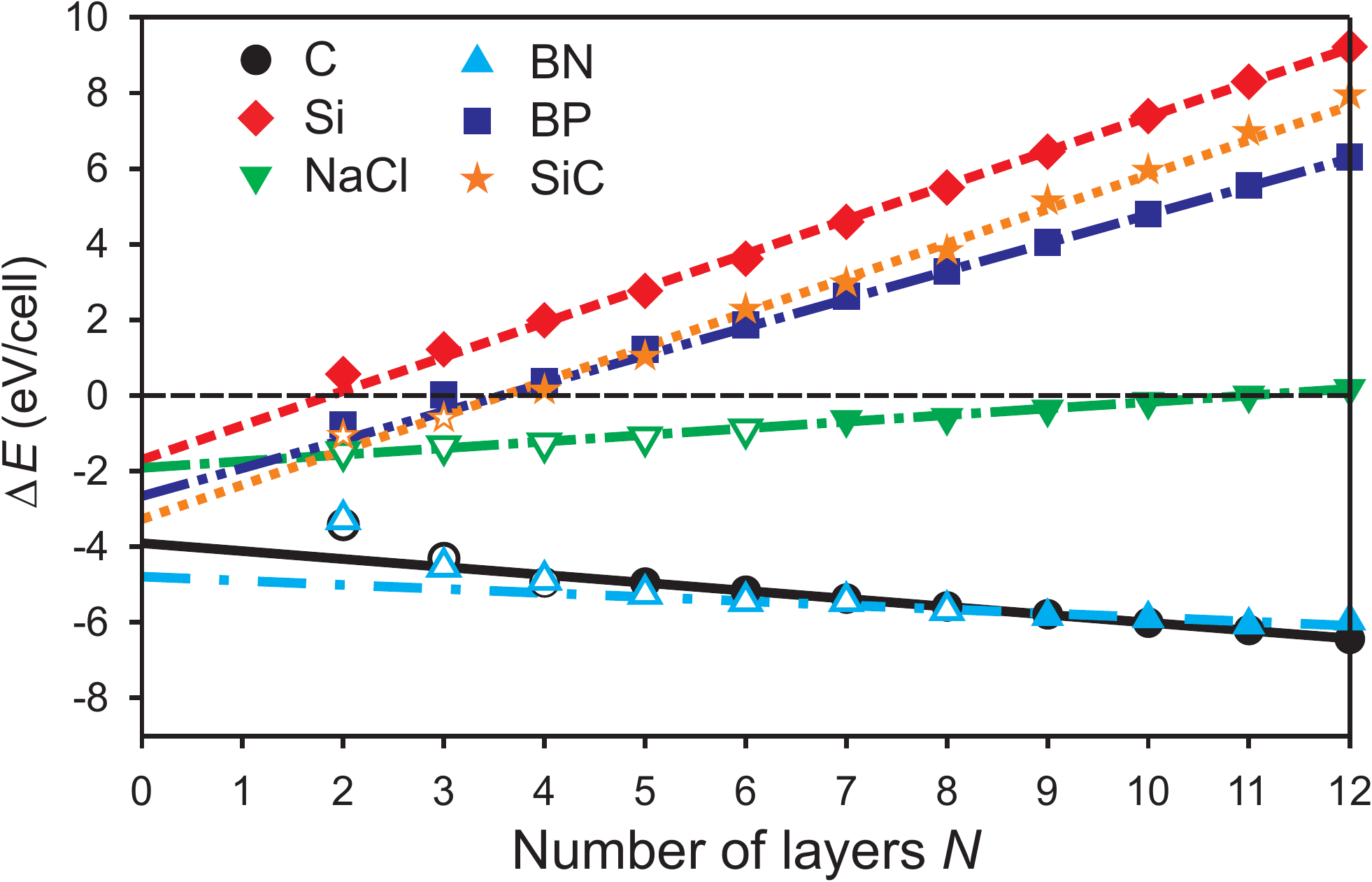}
\caption{(Color online) Cohesive energy difference ${\Delta}E$ of
$N-$layer slabs with cubic and graphitic structure, with
${\Delta}E<0$ indicating energetic preference for graphitization.
Data points are results of DFT calculations. Solid symbols
represent structures with locally stable cubic and graphitic
phase. Open symbols represent structures with an unstable cubic
phase. Lines represent predictions based on
Eq.~(\protect\ref{eq5}) using quantities listed in
Table~\protect\ref{table1}. \label{fig2}}
\end{figure}

The tendency of a system to graphitize can be judged by the
difference of cohesive energies
\begin{equation}
\label{eq1}
{\Delta}E({\rm{bulk}})=E_{cub}({\rm{bulk}})-E_{gra}({\rm{bulk}})
\end{equation}
in the bulk and
\begin{eqnarray}
\label{eq2}
{\Delta}E(N&-&{\rm{layer~slab}})=\\
E_{cub}(N&-&{\rm{layer~slab}})-E_{gra}(N-{\rm{layer~slab}})\nonumber %
\end{eqnarray}
in a free-standing $N-$layer slab. For the sake of consistency, we
subdivide also cubic structures into nominal layers and consider
$N-$layer slabs with the same number of atoms in the cubic and the
graphitic structure. Cohesive energies of bulk and layered
structures are taken per unit cell and are listed in
Table~\ref{table1}. The overbinding of bulk structures, common in
well-converged DFT calculations, does not affect energy
differences. The prevalent energetic preference of the bulk for
the cubic rather than a layered graphitic structure can be
inferred from data in Table~\ref{table1} and is indicated by
${\Delta}E(bulk)>0$.

\begin{table}[t]
\caption{Calculated cohesive and cleavage energies of cubic (cub)
and graphitic (gra) phases of compounds presented in
Fig.~\protect\ref{fig2}. All results are for the (111) cleavage
plane. $N_{c}$ is the critical number of layers for favorable
graphitization according to Fig.~\protect\ref{fig2}.
\label{table1} }
\begin{tabular}{lccccc}
\hline %
& $E_{cub}$(bulk) %
& $E_{gra}$(bulk) %
& $E_{cub}$(cleave) %
& $E_{gra}$(cleave) %
& $N_{c}$ \\
%
& (eV/cell) & (eV/cell) & (eV/cell) & (eV/cell)\\
\hline
C        &  18.18 &  18.39 & 3.91  & $<0.01$ & $\infty$ \\
BN       &  17.47 &  17.57 & 4.83  & $0.04$  & $\infty$ \\
Si       &  10.84 &   9.93 & 2.59  & $0.89$  & 1 \\
SiC      &  14.97 &  14.06 & 3.31  & $0.04$  & 3 \\
BP       &  12.88 &  12.14 & 2.66  & $<0.01$ & 3 \\
NaCl     &   6.78 &   6.60 & 2.34  & $0.42$  & 11 \\
\hline
\end{tabular}
\end{table}

As a counterpart to the bulk results, we plot the dependence of
the slab cohesive energy difference
${\Delta}E(N-{\rm{layer~slab}})$ on the number of layers $N$ for
C, BN, Si, SiC, BN, and for NaCl in Fig.~\ref{fig2}. For
$N\rightarrow\infty$, these results are consistent with the
energetic preference of bulk Si, SiC, BP and NaCl for the cubic
structure, and that of C and BN for the layered graphitic
structure. Our most intriguing result is that ${\Delta}E$ changes
sign in ultrathin slabs of many of the cubic structures,
indicating spontaneous graphitization tendency.

The reason for this behavior is the dominant role of the surface
energy $E({\rm{surface}})$ in the cohesive energy for small values
of $N$. The behavior of ${\Delta}E(N-{\rm{layer~slab}})$ in
Fig.~\ref{fig2} can be explained quantitatively in the following
way. For sufficiently thick slabs, the cohesive energy of an
$N-$layer slab with the cubic structure is given by
\begin{equation}
\label{eq3}
E_{cub}(N-{\rm{layer~slab}}) = %
N E_{cub}({\rm{bulk}}) - E_{cub}({\rm{cleave}})\;,
\end{equation}
where $E_{cub}({\rm{cleave}})=2E_{cub}({\rm{surface}})$ is the
cleavage energy of the bulk cubic crystal or twice the surface
energy per unit cell. Similarly, the cohesive energy of an
$N-$layer slab with the layered graphitic structure is given by
\begin{eqnarray}
\label{eq4}
E_{gra}(N&-&{\rm{layer~slab}}) = \\%
&N& E_{gra}({\rm{bulk}}) - E_{gra}({\rm{cleave}})\nonumber\;,
\end{eqnarray}
where $E_{gra}({\rm{cleave}})$ is the cleavage energy
corresponding to the interlayer interaction per unit cell of the
layered graphitic crystal. Calculated cleavage energies for the
systems of interest in cubic as well as layered graphitic
structures are listed in Table~\ref{table1}. As expected, the
listed values are a small fraction of the bulk cohesive energies
and in general agreement with published data. They ignore
additional energy gain caused by complex surface reconstruction
involving large unit cells, which is a small fraction of the
surface energy\cite{Zangwill} and does not affect our main
predictions.

To get a more quantitative description of the graphitization, we
may combine Eqs.~(\ref{eq2})-(\ref{eq4}) to
%
\begin{eqnarray}
\label{eq5}
{\Delta}E(N&-&{\rm{layer~slab}})=\\
&N& \left[E_{cub}({\rm{bulk}})
-E_{gra}({\rm{bulk}})\right]\nonumber\\
&+&\left[-E_{cub}({\rm{cleave}})+E_{gra}({\rm{cleave}})\right]%
\nonumber\;. %
\end{eqnarray}
The linear dependence of ${\Delta}E$ on the number of layers,
predicted by Eq.~(\ref{eq5}), is reproduced amazingly well in
Fig.~\ref{fig2} down to few layers. Systems with an energetic
preference for the cubic structure in the bulk have a positive
slope, those with a graphitic structure in the bulk have a
negative slope. The reason behind the graphitization of most cubic
structures in our study is the fact that the cleavage or the
surface energy of the graphitic structures is generally lower than
that of cubic structures. This energy difference,
$E_{gra}({\rm{cleave}})-E_{cub}({\rm{cleave}})$, appears as the
intercept of the ordinate in Fig.~\ref{fig2}.

The critical slab thickness for graphitization is determined by
the condition ${\Delta}E(N_c-{\rm{layer~slab}})=0$. In systems
with energetic preference for the cubic phase in the bulk, we
expect graphitization for $N<N_c$ layers and may estimate the
critical value $N_c$ using Eq.~(\ref{eq5}) from
\begin{equation}
\label{eq6}
N_{c}=\frac{E_{cub}({\rm{cleave}})-E_{gra}({\rm{cleave}})}
{E_{cub}({\rm{bulk}})-E_{gra}({\rm{bulk}})} \;. %
\end{equation}
As can be inferred from Fig.~\ref{fig2}, critical slab thicknesses
obtained using the linear extrapolation underlying Eq.~(\ref{eq6})
agree generally very well with $N_c$ values listed in
Table~\ref{table1}, which are based on calculated total energy
differences in finite slabs, where structural changes at the
surface of few-layer systems are considered explicitly. Since only
values $N_c{\agt}2$ indicate graphitization, our finding that
$N_c({\rm{Si}})=1$ agrees with the fact that graphitization of
free-standing silicon slabs to silicene is unfavorable. We should
note that observed silicene layers, which have been stabilized by
strong adhesion to a substrate, are not planar, but buckled,
indicating their
instability and energetic preference for a 3D structure%
\cite{{Feng2012},{Vogt2012},{Lin2012},{Fleurence2012},{Meng2013}}.
The graphitization condition changes to $N>N_c$ in systems, where
the layered graphitic phase is preferred energetically in the
bulk. There, a negative value of $N_c$ obtained using
Eq.~(\ref{eq6}) indicates graphitization for all layer
thicknesses, equivalent to $N_c\rightarrow\infty$ in the
convention used in Table~\ref{table1}.

We need to reemphasize that our results for the graphitization
tendency are given for slabs of the cubic phase terminated by the
(111) surface, which usually has the lowest surface energy. We
expect no qualitative differences when considering other
terminating surfaces.

\begin{figure}[tb]
\includegraphics[width=1.0\columnwidth]{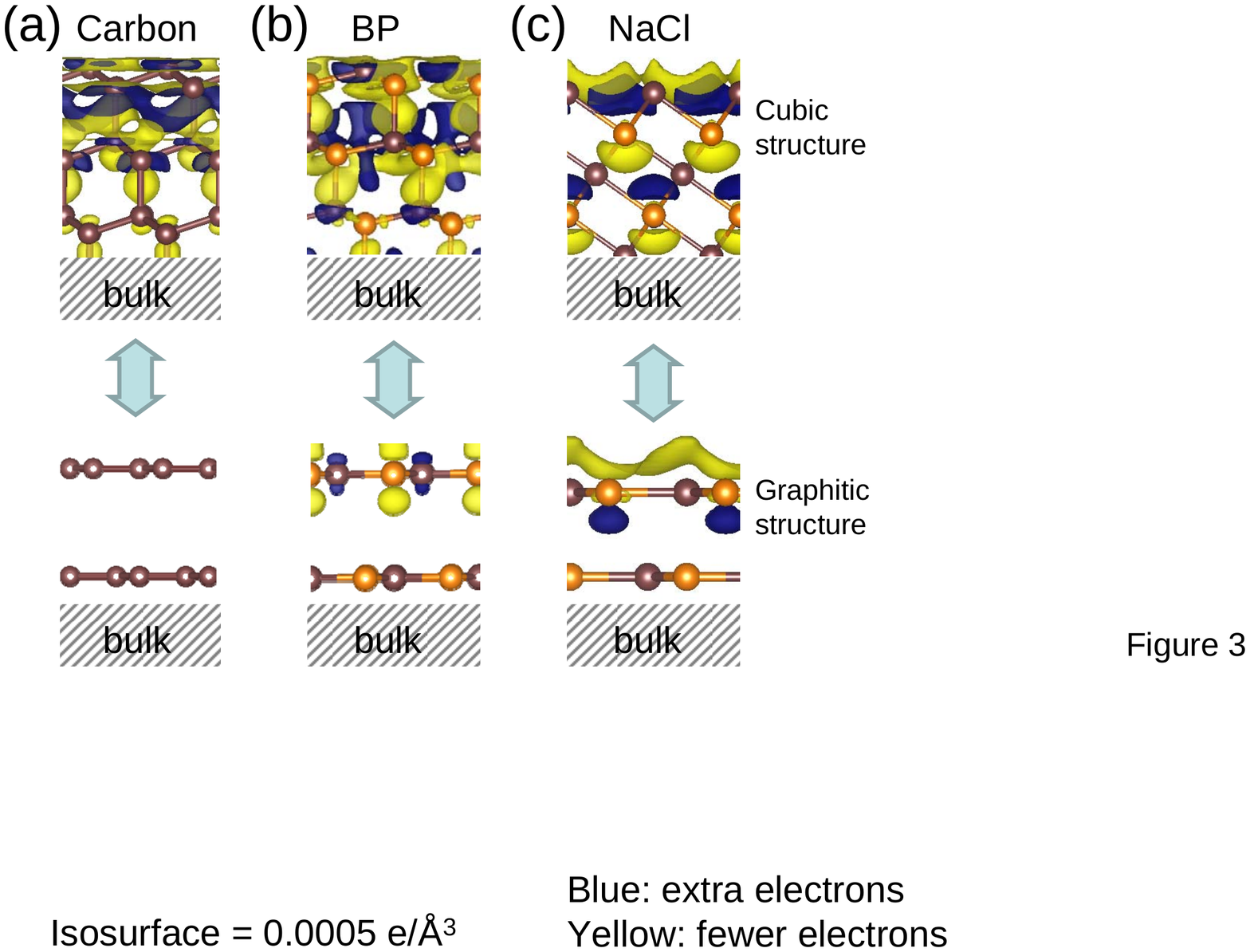}
\caption{(Color online) Total charge density difference
$\Delta\rho$ between the slab and the bulk structure near the
surface of (a) carbon, (b) BP and (c) NaCl in the bulk cubic (top
panels) and the layered graphitic (bottom panels) phase,
superposed with the atomic structure. The isosurface values are
$\Delta\rho_\pm={\pm}5{\times}10^{-4}$~e/{\AA}$^3$ to distinguish
between regions with electron excess, shown by the dark (blue)
isosurfaces, and regions with electron deficit, indicated by the
light (yellow) isosurfaces. \label{fig3}}
\end{figure}

Since the graphitization tendency of ultrathin layers depends
sensitively on the surface or cleavage energy of bulk structures,
we next explore the reasons, why the surface energy of graphitic
structures is generally lower than that of cubic structures. As we
expand later on, the fundamental reason is different in ionic and
in covalent solids. In the latter case, formation of a surface by
cleaving a bulk structure gives rise to unsaturated bonds and a
charge redistribution that is commensurate with the surface
energy. To visualize the degree and the spatial extent of the
charge redistribution, we considered a region of the bulk
structure corresponding to a thick slab, determined the charge
density $\rho({\rm{bulk}})$ in this region, and set $\rho=0$
outside the slab region. Then we truncated the bulk structure to
obtain the geometry of an unrelaxed slab and determined the slab
charge density $\rho({\rm{slab}})$. Finally, we obtained the
charge density difference
$\Delta\rho=\rho({\rm{slab}})-\rho({\rm{bulk}})$ and displayed it
in Fig.~\ref{fig3}.

Inspection of our results in Fig.~\ref{fig3} confirms that charge
redistribution is confined to the surface region and decays to a
vanishing value in the bulk. Comparison between cubic and layered
graphitic structures indicates a significantly lower degree of
charge redistribution in the latter, reflecting our finding that
cleavage and surface energies are lower in layered graphitic than
in cubic structures. This is the case not only in covalent
systems, but -- as seen in Fig.~\ref{fig3}(c) -- also in ionic
systems like NaCl. We find that the degree of charge
redistribution at the surface indeed reflects the relative value
of the cleavage energy as listed in Table~\ref{table1}.

Obviously, the physical origin of energetic stabilization and
charge redistribution at the surface is different in covalent
systems like diamond, in ionic systems such as NaCl, and systems
with covalent and ionic bonding contributions such as BP. In
covalent systems, as mentioned above, surface energy can be
associated with unsaturated dangling bonds. The resulting
significant charge redistribution at the surface of diamond is
clearly visible in the top panel of Fig.~\ref{fig3}(a). The
distribution of $\Delta\rho$ also indicates that the charge flow
in this system is mostly confined to the topmost three layers. In
stark contrast to these findings, the charge redistribution caused
by cleaving graphite, displayed in the bottom panel of
Fig.~\ref{fig3}(a), is significantly smaller. The absence of
contours in this figure indicates that the charge density
difference between the bulk and the surface lies below
$5{\times}10^{-4}$~e/{\AA}$^3$.

The ionic nature of bonding gives rise to a large surface dipole
at the (111) surface of the cubic NaCl structure, which is the
origin of the large surface energy\cite{FreemanPRL06}. As seen in
the top panel of Fig.~\ref{fig3}(c), the charge redistribution at
the surface of NaCl is significant and, due to lack of screening,
involves more layers than in covalent materials. As an alternative
to the bulk cubic structure, we can imagine arranging Na and Cl
atoms in charge neutral honeycomb layers that would form a layered
structure. As pointed out earlier\cite{FreemanPRL06}, the major
energetic benefit of the layered graphitic structure of NaCl
results from removing the surface dipole component normal to the
surface, thus reducing the electrostatic energy penalty. Our
results for $\Delta\rho$ in the bottom panel of Fig.~\ref{fig3}(c)
indicate a much smaller degree of charge redistribution in the
graphitic layered structure and also a stronger confinement to the
narrow surface region, explaining the significant difference in
the cleavage or surface energy between the layered graphitic and
the cubic structure, listed in Table~\ref{table1}.

Our results for boron phosphide, shown in Fig.~\ref{fig3}(b),
indicate similarities with covalent and ionic systems in
Figs.~\ref{fig3}(a) and \ref{fig3}(c). For one, cleavage of the
cubic phase affects the charge distribution in more surface layers
than in purely covalent crystals. We also observe a noticeable
charge redistribution upon cleavage of the layered graphitic
phase, which is mostly confined to the topmost layer.

The results presented above have focussed on energy differences
between two structural phases, but say little about the local
stability of these structures or about a way to synthesize them.
As mentioned before, the open symbols in the ${\Delta}E<0$ region
of Fig.~\ref{fig2} indicate instability of the cubic phase and its
spontaneous conversion to a layered graphitic phase. For other
systems, we find both phases to be locally stable, implying an
activation barrier for the conversion. Such energy barriers may be
factual or may result from unit cell size and symmetry constraints
imposed in our calculations. In an infinite slab with no such
constraints, possibly aided by the presence of defects, such
activation barriers may be strongly suppressed or even vanish,
providing an energetically viable pathway for the structural
change. The energetics of the conversion of NaCl from the bulk
cubic to the layered graphitic structure is discussed in the
Supplemental Material\cite{SM-cubgra14}.

The geometry of samples formed by Chemical Vapor Deposition (CVD)
closely resembles the constrained optimized geometry described
here. Recent success achieved in CVD synthesis of layered
structures including graphene\cite{CVDgraphene09} and hexagonal
boron nitride\cite{BN-CVD2010} indicates that this approach may
also be useful to form ultrathin slabs of other compounds with a
layered graphitic structure. An important criterion for the
selection of the deposition substrate is the requirement that the
substrate-adlayer interface energy should not penalize
energetically the formation of a graphitic structure.
Free-standing few-layer slabs could then be obtained by mechanical
transfer of the deposited structure.


In summary, we performed {\em ab initio} density functional
calculations that indicate a general graphitization tendency in
ultrathin slabs of cubic compounds. We found that an energy
preference for layered honeycomb rather than cubic structures in
ultrathin slabs is rather common, extending from diamond and boron
nitride to less obvious cubic structures including silicon
carbide, boron phosphide and rocksalt. Whereas the bulk of such
compounds shows an energy preference for cubic rather than layered
atomic arrangements, the surface energy of systems with honeycomb
layers is commonly lower than that of their cubic counterparts.
Whether the type of crystal bonding is purely covalent, purely
ionic, or a combination of the two, the optimum structure of a
slab results from an energy competition between the energy
preference for a honeycomb structure at the surface and for a
cubic atomic arrangement in the bulk. We determined the critical
slab thicknesses for a range of systems, below which a spontaneous
conversion from a cubic to a layered graphitic structure occurs,
driven by surface energy reduction in surface-dominated
structures. Finally, we believe that graphitization of ultrathin
layers is a rather general phenomenon that is not limited to
systems in this study and expect that it will provide a viable
route to a bottom-up synthesis of few-layer compounds by CVD.

\begin{acknowledgements}
This study has been supported by the National Science Foundation
Cooperative Agreement \#EEC-0832785, titled ``NSEC: Center for
High-rate Nanomanufacturing''. Computational resources have been
provided by the Michigan State University High Performance
Computing Center and by the Supercomputing Center of the Lomonosov
Moscow State University. AGK was supported by a Scholarship from
the President of Russia for young scientists and PhD students
(competition SP-2013). PBS acknowledges the hospitality of
Michigan State University, where this research was performed, and
additional support by the FAEMCAR Grant No. FP7-PEOPLE-2012-IRSES.
\end{acknowledgements}



\begin{thebibliography}{23}
\bibitem [{2D-()}]{2D-constraints}%
  \BibitemOpen
  \href@noop {} {}\bibinfo {note} {The growth mechanism of films on selected
  substrates by Chemical Vapor Deposition (CVD) differs fundamentally from the
  growth of nanostructures in free space. Whereas aggregation of carbon atoms
  leads to the formation of fullerenes in free space, layered graphitic
  structures are formed by CVD on a substrate. Our 2D constrained optimization
  addresses the latter process.}\BibitemShut {Stop}%
\end{thebibliography}

%

\end{document}